\documentclass[letterpaper, 10 pt, conference]{ieeeconf}  

\IEEEoverridecommandlockouts                              

\usepackage{graphics} 
\usepackage{graphicx}
\usepackage{epsfig} 
\usepackage{times} 
\usepackage{balance}

\usepackage{color}
\usepackage{epstopdf}
\usepackage[utf8]{inputenc}
\usepackage{amsmath}
\usepackage{amssymb}
\usepackage{makecell}
\usepackage{booktabs}
\usepackage{mathrsfs}
\usepackage{theorem}
\usepackage{algorithm}
\usepackage{algorithmic}
\usepackage{tabularx}
\usepackage{indentfirst}
\usepackage{float}

\newtheorem{theorem}{Theorem}

\newtheorem{remark}{Remark}

\usepackage{cleveref}

\title{\LARGE \bf  Distributed Consensus Optimization  with Consensus ALADIN
}
\author{Xu Du, Jingzhe Wang$^*$
    \thanks{$^*$ Corresponding author.}
\thanks{Xu Du, is with the Artificial Intelligence Thrust of the Information Hub, The Hong Kong University of Science and Technology (Guangzhou), Guangzhou, China.  E-mail: \texttt{duxu@hnas.ac.cn}.}
	\thanks{Jingzhe Wang is with the Department of Informatics \& Networked
		Systems, School of Computing and Information, University of Pittsburgh,
		Pittsburgh, PA, USA.  E-mail: \texttt{jiw148@pitt.edu}.}
}

\begin{document}
	
	\maketitle
	\thispagestyle{empty}
	\pagestyle{empty}

	\begin{abstract}
		The paper proposes the Consensus Augmented Lagrange Alternating Direction Inexact Newton (Consensus ALADIN) algorithm, a novel approach for solving distributed consensus optimization problems (DC).  Consensus ALADIN allows each agent to independently solve its own nonlinear programming problem while coordinating with other agents by solving a consensus quadratic programming (QP) problem.  Building on this, we propose Broyden–Fletcher–Goldfarb–Shanno (BFGS) Consensus ALADIN, a communication-and-computation-efficient Consensus ALADIN. BFGS Consensus ALADIN improves communication efficiency through BFGS approximation techniques and enhances computational efficiency by deriving a closed form
for the consensus QP problem.  Additionally, by replacing the BFGS approximation with a scaled identity matrix, we develop Reduced Consensus ALADIN, a more computationally efficient variant.  We establish the convergence theory for Consensus ALADIN and demonstrate its effectiveness through application to a non-convex sensor allocation problem.

	\end{abstract}

	\section{Introduction}\label{sec: Intro}
	Distributed optimization problems with affine coupled constraints (DO) are generally formulated in the fashion of mathematical programming, where $N$
	separable objectives are linearly coupled by $m$ equality constraints. In DO, 
	$f_i: \mathbb R^{n_i}\rightarrow \mathbb R$ denotes the closed proper local objective function of each agent $i$, where $i\in \mathcal I=\{1, 2, ..., N\}$. 
	Formally, DO can be described as follows \cite{Houska2021},
	\begin{equation}\label{eq: DOPT_G}
		\begin{split}
			\min_{x_i\in \mathbb R^{n_i}, \forall i \in \mathcal I}& \quad \mathop{\sum}_{i=1}^{N}  f_i(x_i)\\
			\mathrm{s.t.}\;\quad\;& \quad\mathop{\sum}_{i=1}^{N} A_ix_i=b. 
		\end{split}
	\end{equation}
	Here, the coupling matrices $A_i\in \mathbb R^{n_i\times m}$ and the coupling parameter $b\in \mathbb R^{m}$ are given. The dimensions $n_i$s of local variables $x_i$s may differ.
		In recent years, DO has gained significant attention due to its successful applications to problems in signal processing \cite{boyd2011distributed},  model predictive control \cite{Houska2021,chanfreut2023aladin} and power grids \cite{Engelmann2019, Timm2024}. 
 
 Distributed consensus  optimization (DC) problems, which are a special case of DO, are formulated as follows:
	\begin{equation}\label{eq: DC}
		\begin{split} 
			\min_{x_i, z\in \mathbb R^n,\forall i \in \mathcal I}& \quad\sum_{i=1}^{N}f_i(x_i) \\ \quad\mathrm{s.t.}\quad\;\;&\quad x_i = z \;|\lambda_i.
		\end{split}
	\end{equation} 
 Here, $f_i: \mathbb R^{n}\rightarrow \mathbb R$ denotes closed and proper local objective function of agent $i$, and $\lambda_i$ represents the dual variable associated with agent $i$. We introduce $x$ and $\lambda$ to stack $x_i$ and $\lambda_i,\forall i$ as vectors respectively. The notation $(a\vert b)$ denotes the relation between the constraint $(a\in \mathbb R^n)$ and the corresponding dual variable $(b\in \mathbb R^n)$.
	The main difference between DO and DC problems is that DC includes a global variable $z$ to which all private variables $x_i$s must converge. 
	Applications of DC include, but are not limited to, federated learning  (FL) \cite{ioffe2015batch} and sparse signal processing \cite{boyd2011distributed}.
	As a milestone in DC research, \cite{shi2014linear} shows that Consensus Alternating Direction Method of Multiplier (Consensus ADMM) \cite{boyd2011distributed} achieves a linear convergence rate for DC problems that are strongly convex under some assumptions. We refer \cite{yang2022survey} as a survey paper for more details. 
	It is worth noting that, similar to Consensus ADMM, other algorithms such as Decentralized Gradient Descent (DGD) \cite{yuan2016convergence}, Exact First-Order Algorithm (EXTRA) \cite{shi2015extra} provide convergence guarantees for decentralized convex problems in the the study of DC. However, a detailed discussion of decentralized optimization algorithms is beyond the scope of this paper.
	
	In this paper, we focus on a recent algorithm, named Augmented Lagrangian based Alternating Direction Inexact Newton method (ALADIN) \cite{houska2016augmented}. ALADIN was originally developed for solving DO \eqref{eq: DOPT_G} problems and can be viewed as a combination of ADMM and sequential quadratic programming (SQP) \cite{Nocedal2006}. ALADIN guarantees local convergence for non-convex DO problems and global convergence for convex DO problems.
	To date, ALADIN has inspired several elegant successors \cite{Engelmann2019,Du2019,Houska2021,engelmann2020decomposition} and has demonstrated effectiveness in many applications \cite{Jiangwireless,Engelmann2019,Du2019,chanfreut2023aladin}. 
		Despite these advancements, it remains an open question whether ALADIN can solve Problem \eqref{eq: DC} efficiently. 
	This naturally leads us to ask:
	\textit{Can we propose novel variants of ALADIN to solve DC~(Problem~\eqref{eq: DC}) efficiently while providing formal convergence guarantees?}

\textbf{Contribution:} In this paper, we first propose Consensus ALADIN, a novel algorithmic structure that aligns the ALADIN approach with the structure of Problem~\eqref{eq: DC}. Consensus ALADIN enables each agent to solve its own nonlinear programming problem while coordinating with other agents by solving a consensus quadratic
programming (QP) problem. Building on this, we propose Broyden–Fletcher–Goldfarb–Shanno (BFGS) Consensus ALADIN, a communication-and-computation-efficient variant of Consensus ALADIN. In BFGS Consensus ALADIN, we reconstruct the Hessian matrix of the sub-problems using the BFGS approximation technique, thereby avoiding the need to transmit the full Hessian matrices. This enhances communication efficiency compared to directly transmitting the matrix in Consensus ALADIN. Furthermore, BFGS Consensus ALADIN leverages a closed-form solution for the consensus QP, 
resulting in enhanced computational efficiency. Additionally, based on BFGS Consensus ALADIN, we propose Reduced Consensus ALADIN, which replaces the Hessian matrix with a scaled identity matrix to further enhance computational efficiency. We establish convergence theory of  Consensus ALADIN for Problem~\eqref{eq: DC}. Specifically, we provide a global convergence theory for convex DC problems and a local convergence theory for non-convex DC problems. Our numerical experiments on a non-convex sensor allocation problem demonstrate the effectiveness of our proposed algorithms.

	\textbf{Organization:}~The paper is structured as follows: Section \ref{sec: Preliminaries} provides the preliminaries of ALADIN and its limitation for solving \eqref{eq: DC}.  In Section~\ref{sec: Consensus ALADIN},
	we introduce Consensus ALADIN, a novel family of algorithms designed to address Problem \eqref{eq: DC}. Section \ref{sec: convergence} provides the convergence analysis of Consensus ALADIN.
	In Section~\ref{sec: numerical}, we perform numerical experiments. Section \ref{sec: summary} concludes this paper.
	
	\textbf{Notations:}
    In this paper, $(\cdot)^-$ denotes the value from the previous iteration, while $(\cdot)^+$ represents the value from the current iteration. For ease of expression, $(\cdot)^k$ indicates the value of $(\cdot)$ at the $k$-th iteration for the given algorithms. Additionally, $\|\cdot\|_{M}= \sqrt{(\cdot)^\top M(\cdot)}$ when  $M\succ 0$.

	\section{Background and Motivation}\label{sec: Preliminaries}
 In this section, we first review the basics of ALADIN in Section~\ref{sec: basics of ALADIN}, followed by a motivating example in Section~\ref{sec: disadvantages}.
	\subsection{Preliminaries of ALADIN}\label{sec: basics of ALADIN}
	At a high level, ALADIN allows each agent to independently address its nonlinear programming (NLP) problem while communicating with other agents by solving a coupled QP problem.
	The detailed algorithmic structure of ALADIN is given in Algorithm \ref{alg:ALADIN}.
	\begin{algorithm}[ht]
		\small
		\caption{Standard ALADIN}
		\textbf{Initialization:} Initial guess of primal and dual variables $(\epsilon_i,\mu)$.  \\
		\textbf{Repeat:}
		\begin{enumerate}
			\item Parallelly solve local NLP:
			\begin{equation}\label{ALADIN-step1}
				{x_i}^+=\mathop{\arg\min}_{x_i} f_i(x_i)+\mu^\top A_i x_i+\frac{\rho}{2}\|x_i-\epsilon_i\|^2.
			\end{equation}
			\item Evaluate and upload the Hessian approximation $B_{i}\succ 0$ and the gradient $g_{i}$ at ${x_i}^+$:
			\begin{equation}\label{eq: ALADIN upload}
				\left\{
				\begin{array}{l}
					\begin{split}
						B_{i}\approx &\nabla^2 f_i(x_i^+), \\
						g_{i}=&\rho\left(\epsilon_i-x_i^+\right)-A_i^\top \mu.
					\end{split}
				\end{array}
				\right.
			\end{equation}
			\item Solve the coupled QP on master side with the uploaded $x_i^+,\;B_i,\;g_i$ from each agent:
			\begin{equation}\label{eq: ALADIN-coupled QP}
				\begin{split}
					\mathop{\mathrm{\min}}_{ \Delta x_i,\forall i \in \mathcal I}& \quad \mathop{\sum}_{i=1}^{N} \frac{1}{2}\Delta x_i^\top B_{i} \Delta x_i+g_{i}^\top \Delta x_i \\
					\mathrm{s.t.} \;\;\;&\quad\mathop{\sum}_{i=1}^{N} A_i(x_i^++\Delta x_i)=b\; |{\mu}^+	.
				\end{split}
			\end{equation}
			
			\item Download:
			\begin{equation}\label{eq: download}
				\left\{
				\begin{array}{l}
					\begin{split}
						\mu& \leftarrow {\mu}^+,\\
						\epsilon_{i}&\leftarrow x_i^++\Delta x_i.
					\end{split}
				\end{array}
				\right.
			\end{equation} 
		\end{enumerate}
		\label{alg:ALADIN}
	\end{algorithm}
	
	In Algorithm~\ref{alg:ALADIN}, the initial step involves each agent updating its local minimizer to $x_i^+$. Given $x_i^+$, each agent then approximates the positive definite Hessian matrices $B_i$s, and evaluates the gradients $g_i$s of $f_i$s, as per \eqref{eq: ALADIN upload} in Step 2).
	Upon receiving $B_i$ and $g_i$ from all agents, the master solves a large-scale convex coupled QP \eqref{eq: ALADIN-coupled QP}.
	The solution to this QP \eqref{eq: ALADIN-coupled QP} includes the updated primal variables $\left(x_i^+ + \Delta x_i\right)$s and dual variable $\mu$ in Step 3). Finally, the master communicates the updated primal and dual variables back to each agent, thereby concluding the current iteration in Step 4).
 
	The convergence behavior of ALADIN can be summarized as follows: For non-convex DO problems \cite{houska2016augmented}, Algorithm~\ref{alg:ALADIN} has local convergence guarantees if the linearly independent constraint qualification (LICQ) and second order sufficient condition (SOSC) of Problem \eqref{eq: DOPT_G} are locally satisfied. In the case of convex problems, ALADIN has global convergence guarantees without requiring the smoothness of the objectives if the strong duality of Problem \eqref{eq: DOPT_G} holds \cite{Houska2021}. For more details, we refer interested readers to \cite{Houska2021}.
	\subsection{Limitations of ALADIN for Solving DC}\label{sec: disadvantages}
	In this subsection, we analyze the limitations of ALADIN when directly applied to solve DC, which motivates us to present our new algorithm family.
 
	In general, there are two straightforward approaches that may adapt ALADIN for DC. 
	Specifically, the \textbf{first} approach applies ALADIN directly to \eqref{eq: DC}.
	Such an approach, however, is clearly flawed. The primary issue is that ALADIN, originally tailored for DO problems \eqref{eq: DOPT_G}, requires each variable to have a strictly positive Hessian approximation matrix $B_i\succ0$ corresponding to its independent objective function. Problem \eqref{eq: DC} presents a challenge, as the global variable $z$ is not linked to an independent objective function, thereby violating a fundamental principle of ALADIN. 
 
	The \textbf{second} way involves a more common approach to solving DC as mentioned in \cite[Section 12]{ALADIN2023learning}, which suggests reformulating \eqref{eq: DC} to align with standard DO fashion and then adopting DO algorithms. The reformulation of \eqref{eq: DC} is elaborated as follows,
	\begin{equation}
		\label{eq: DOPT_G2}\small
		\begin{split}
			\min_{x_i\in \mathbb R^{n}, \forall i \in \mathcal I}& \quad   f_1(x_1)+f_2(x_2)+\cdots+f_N(x_N)\\
			\mathrm{s.t.}\;\quad\;&  \quad\begin{matrix}
				\;\;\; x_1  &-x_2 & & & &=0 \\
				&\;\;\; x_2 &-x_3 & & &=0\\
				&   \vdots & \vdots& &\vdots & \;\;\;\;\vdots\\
				&    &  & x_{N-1}
				& -x_N&=0\\
				-x_1 &    &  & 
				& \;\;\;x_N&\;=0.\\
			\end{matrix}
		\end{split}
	\end{equation}
	The linear constraints in  Problem \eqref{eq: DOPT_G2} can be described as $\mathop{\sum}_{i=1}^{N} A_ix_i=0,$
	where $A_1 = \left(I,
	0,
	\cdots,
	0,
	-I \right)^\top$, $A_2=\left(-I,
	I,
	0,
	\cdots,
	0 \right)^\top$,$\cdots$, $A_N=\left(0,
	\cdots,0,
	-I,
	I \right)^\top$.
	Here $I$ denotes the identity matrix with proper dimension. We denote the set of such vectors as $A=\{A_i\}_{i \in \mathcal{I}}$.
	However, this approach has two limitations. First, the selection of $\{A_i\}$ is not unique, and these linear coupling matrices $A_i$s can be challenging to implement in practice. 
   Second, employing Algorithm \ref{alg:ALADIN} to solve \eqref{eq: DOPT_G2} is inefficient in terms of both communication and computational. Specifically, the communication overhead arises from transmitting local Hessian matrices $\nabla^2 f_i(x_i^+)$ and gradients $\nabla f(x_i^+)$ of the local NLPs at each iteration, as detailed in \eqref{eq: ALADIN upload}. Additionally, coordination requires solving a large-scale coupled QP problem, as outlined in \eqref{eq: ALADIN-coupled QP}, which results in significant computational overhead.


	\section{Consensus Augmented Lagrange Alternating Direction Inexact Newton Method}\label{sec: Consensus ALADIN}
	
	In Subsection \ref{sec: consensus QP},  we present the consensus QP and our proposed Consensus ALADIN. 
	Subsection~\ref{sec: Algorithms} details  Consensus ALADIN, consisting of two variants, namely BFGS  Consensus ALADIN and Reduced Consensus ALADIN.
	
	\subsection{Consensus ALADIN: From Coupled QP to Consensus QP}\label{sec: consensus QP}
 Recall from Subsection \ref{sec: disadvantages} that ALADIN faces challenges in solving Problem \eqref{eq: DC} due to the complexity of communication and computation in Algorithm \ref{alg:ALADIN}, as well as the inefficient implementation of coupling equality constraints in Equation \eqref{eq: ALADIN-coupled QP} and \eqref{eq: DOPT_G2}.
 Drawing inspiration from Consensus ADMM \cite{boyd2011distributed}, our first step towards making ALADIN viable in \eqref{eq: DC}  involves reconstructing Equation \eqref{eq: ALADIN-coupled QP}. This leads to the formulation of our convex consensus QP, formally shown as follows,
	\begin{equation}\label{eq: consensus QP}
		\begin{split}
			\mathop{\mathrm{\min}}_{ \Delta x_i, z\in \mathbb R^n,\forall i \in \mathcal I}& \quad \mathop{\sum}_{i=1}^{N} \left(\frac{1}{2}\Delta x_i^\top B_i \Delta x_i+g_{i}^\top \Delta x_i\right) \\
			\mathrm{s.t.}\qquad \;&\quad x_i^++\Delta x_i=z\; |\lambda_{i}.
		\end{split}
	\end{equation}
	Notably, similar to Consensus ADMM, \eqref{eq: consensus QP} introduces a global variable $z$ and couples all $x_i$s to $z$.
	Note that,  
	Equation \eqref{eq: consensus QP} differs from the original formulation in Equation \eqref{eq: ALADIN-coupled QP} by incorporating one global primal variable $z$ and $N$ dual variables $\lambda_i$s.
	
	After replacing Equation \eqref{eq: ALADIN-coupled QP} with Equation \eqref{eq: consensus QP}, Algorithm \ref{alg:ALADIN} is now capable of operating in a consensus fashion. We name this Consensus ALADIN.

	\subsection{Algorithms}\label{sec: Algorithms}
	\subsubsection{\textbf{BFGS  Consensus ALADIN}}
	To improve the communication and computation efficiency of Consensus ALADIN,  we detail BFGS  Consensus ALADIN (a variant of Consensus ALADIN) in Algorithm \ref{alg:BFGS ALADIN2}.
	\begin{algorithm}[h]
		\caption{BFGS   Consensus  ALADIN}
		\textbf{Initialization:} choose $\rho>0$, initial guess $(\lambda_i,z,B_i=\rho I)$.\\
		\textbf{Repeat:}
		\begin{enumerate}
			\item Each agent optimizes its own variable $x_i$ locally and transmits it to the master
			\begin{equation}\label{eq: NLP}
				{x_i}^+=\mathop{\arg\min}_{x_i} f_i(x_i)+\lambda_i^\top  x_i+\frac{\rho}{2}\|x_i-z\|^2
			\end{equation}
			with
			\begin{equation}\label{eq: lam update}
				\lambda_{i}=B_i^-(x_i^--z)-g_i.
			\end{equation}

			\item  Operations on the master side:\\
			a) Recover the gradient and BFGS Hessian  from each ${x_i}^+$:
			\begin{equation}\label{eq: subgradient}
				\left\{
				\begin{split}
					g_i=&\;\rho(z-x_i^+)-\lambda_i& \text{ (local gradient evaluation)},\\
					s_i=&\;x_i^+-x_i^- &\text{(difference of local variables)},\\
					y_i=&\; g_i-g_i^-& \text{(difference of local gradient).}
				\end{split}
				\right.
			\end{equation}
			b) Modify the local gradient with $y_i=y_i + \theta (B_is_i-y_i)$ where
			\[\theta= \frac{0.2(s_i)^\top B_i s_i-(s_i)^\top y_i}{(s_i)^\top B_i s_i-(s_i)^\top y_i}\]  if  $(y_i)^\top s_i\leq\frac{1}{5} (s_i)^\top B_i s_i$.
			
			c) \text{BFGS Hessian approximation evaluation (optionally)\footnotemark{}:}
			\begin{equation}\label{eq: BFGS}
				B_i=\;B_i^--\frac{B_i^- s_i s_i^\top B_i^-}{s_i^\top B_i^- s_i}+\frac{y_iy_i^\top}{s_i^\top y_i}.
			\end{equation}
			d) The master updates the global variable $z$ with  \begin{equation}\label{eq: reduced QP}
				\begin{split}
					z=\left(\sum_{i=1}^N  B_i \right)^{-1}  \sum_{i=1}^N \left(B_ix_i^+ -    g_i   \right).
				\end{split}
			\end{equation}
		\end{enumerate}
		\label{alg:BFGS ALADIN2}
	\end{algorithm}
	\footnotetext[1]{One can optionally update the the Hessian matrices at iteration $k$ with $K\in \mathbb N^+$, if $\text{log}_K(k)\in \mathbb{N}^+$ \cite[Section III-C]{Shi2022}. 
	}	
	
	Similar as Algorithm \ref{alg:ALADIN}, in the first step, each agent updates it's own local variable $x_i^+$  with a recovered $\lambda_i^-$ and uploads it to the master. Later, the master recovers the agents' gradient and Hessian  approximation in the second step by using the damped BFGS technique (\cite[Page 537]{Nocedal2006}). Finally, the master updates the global variable $z^+$ with the Equation \eqref{eq: reduced QP}  and broadcasts it to the agents. Repeat these steps until $z$ converges. Here, \eqref{eq: lam update} and \eqref{eq: reduced QP} are derived from the Karush-Kuhn-Tucker (KKT) system of Problem \eqref{eq: consensus QP}.
	
	\begin{remark}
		Note that, similar to Step 3) in Algorithm \ref{alg:ALADIN}, \eqref{eq: consensus QP}
		also requires the primal variable $x_i^+\in \mathbb R^n$, the gradient $g_i\in \mathbb R^n$, and the local Hessian approximation $B_i\in \mathbb R^{n\times n}$. 
		In total, this requires transmitting $\left(2Nn + Nn^2\right)$ floats. Specifically, when $N$ is fixed, transmitting $g_i$ requires $Nn$ floats, while transmitting $B_i$ requires $Nn^2$ floats. The transmission of these two components is the main factor affecting communication efficiency during the upload phase. Therefore, Step 2) of Algorithm \ref{alg:BFGS ALADIN2} improves upload communication efficiency by avoiding the direct transmission of the gradient and Hessian approximation \eqref{eq: ALADIN upload}. Taking advantage of our design, BFGS Consensus ALADIN only requires transmitting $Nn$ floats on the upload phase.
	\end{remark} 
	
	\begin{remark}
		Solving  the KKT system of Problem \eqref{eq: consensus QP} finds the inversion of a KKT matrix with dimension  $\left((2N+1)n\times(2N+1)n\right)$. To alleviate this issue, we design a new method that updates the global variable $z$ directly without updating the $\lambda_i$s and $\Delta x_i$s, avoiding solving \eqref{eq: consensus QP} directly. The optimal solution of global variable $z$ in Equation \eqref{eq: consensus QP} has a closed form as Equation \eqref{eq: reduced QP} and only relies on deriving the inversion of an $n\times n$ matrix. 
	\end{remark}

	\subsubsection{\textbf{Reduced Consensus ALADIN}}
	To further improve the computational efficiency of \eqref{eq: reduced QP}, we next introduce our proposed Reduced Consensus ALADIN, as an extension of BFGS  Consensus ALADIN. Our Reduced Consensus ALADIN differs from BFGS Consensus ALADIN in that the global variable $z$ is updated based on  the following equation,
	\begin{equation}\label{eq: reduced QP2}
		z=\frac{1}{N} \sum_{i=1}^{N} \left(x_i^+- \frac{g_i}{\rho}\right). 
	\end{equation}

 For Reduced Consensus ALADIN, the updates of $x_i^{+}$, $\lambda_i$, $g_i$ and $z$ in Algorithm \ref{alg:BFGS ALADIN2} can be easily replaced with those defined by the following equation,
	\begin{equation}\label{eq: RCA}
		\left\{ \begin{split}
			&{x_i}^+=\mathop{\arg\min}_{x_i}f_i(x_i)+\lambda_i^\top  x_i+\frac{\rho}{2}\|x_i-z\|^2,\\
			&g_i=\rho(z-x_i^+)-\lambda_i,\\
			&z^+=	\frac{1}{N} \sum_{i=1}^{N}\left(x_i^+- \frac{g_i}{\rho}\right),\\
			&\lambda_{i}^+=\rho(x_i^+-z^+)-g_i.
		\end{split}\right.
	\end{equation}
	Note that, both two algorithms (Algorithm \ref{alg:BFGS ALADIN2} and Equation  \eqref{eq: RCA}) belong to the class of Consensus ALADIN. 
	As a supplement, the difference between the Reduced Consensus ALADIN and Consensus ADMM can be found in the appendix.
	%

	\section{Convergence Analysis}\label{sec: local convergence}\label{sec: convergence}
	In this section, we present the convergence analysis for Consensus ALADIN. Specifically, Section~\ref{sec: global} develops the global convergence theory for convex DC problems, while Section~\ref{sec: local} provides the local convergence analysis for non-convex DC problems. Our results for Consensus ALADIN are also applicable to BFGS Consensus ALADIN and Reduced Consensus ALADIN.

	\subsection{Global Convergence Analysis}\label{sec: global}
	We assume that the  $f_i$s of \eqref{eq: DC} are closed, proper, and strictly convex. 
	In the global convergence theory of Consensus ALADIN, the sub-problems are updated as,
			\begin{equation}\label{eq: x update2}
				{x_i}^+=\mathop{\arg\min}_{x_i} f_i(x_i)+\lambda_i^\top  x_i+\frac{1}{2}\|x_i-z\|_{B_i}^2
			\end{equation}
   instead of \eqref{eq: NLP}.
			For ease of analysis, we temporarily assume that the matrices $B_i\succ 0$  are constant.

	For establishing the global convergence theory of  Consensus ALADIN,		
	we introduce the following \emph{energy function} \cite{ling2015dlm} (also called Lyapunov function in \cite[Appendix A]{boyd2011distributed} and \cite{yang2022proximal}) with the unique optimal solution $(z^*, \lambda^*)$,
	\begin{equation}\label{eq: LYA}
		\mathscr{L}(z,\lambda)=	 \sum_{i=1}^{N}\left(\left\|\lambda_i-\lambda_i^*\right \|_{B_i^{-1}}^2 +   \left\|z-z^*\right\|_{B_i}^2\right).
	\end{equation}
	Note that the choice of energy function is not unique \cite{nonlinear}. Next, we will establish the global convergence of Consensus ALADIN by demonstrating the monotonic decrease of the energy function \eqref{eq: LYA}.


	\begin{theorem}\label{The: 1}
		Let the local objective  $f_i$s of Problem~\eqref{eq: DC} be closed, proper,  strictly convex. Let the strong duality of Problem~\eqref{eq: DC} hold. Let matrices $B_i\succ 0$s  be constant.  Let $x_i^*=z^*$ denote the primal solution and $\lambda^*$ denotes the dual solution of Problem~\eqref{eq: DC}, then
		\begin{equation}
			\mathscr{L}(z^+,\lambda^+)\leq \mathscr{L}(z,\lambda)-4\Pi\left(\sum^N_{i=1}\left\|x_i^+-z^*\right\|\right)
		\end{equation}
		is satisfied with the iterations of Algorithm \ref{alg:BFGS ALADIN2}.
		Here, $\Pi$ is a class $\mathcal K$ function \cite{nonlinear}.		
	\end{theorem} 
	\textbf{Proof:}
	First, we introduce the auxiliary functions \begin{equation}\label{eq: aux}
		\left\{
		\begin{split}
			\tilde{f}_i(\chi_i)&=f_i(\chi_i)+(\lambda_i+B_i(x_i^+-z))^\top \chi_i,\\
			G(\chi)&=\sum_{i=1}^{N}f_i(\chi_i)+\sum_{i=1}^{N}(\chi_i-z^*)^\top\lambda_i^*,
		\end{split}\right.
	\end{equation}
    where $\chi_i\in \mathbb R^n$.
	Due to the strict convexity of both functions in \eqref{eq: aux}, the following equation holds,
	\begin{equation}\label{eq: aux_ineq}
		\left\{
		\begin{split}
			\tilde{f}_i(x_i^+)&\leq\tilde{f}_i(z^*)-\tilde{\Pi}_i\left(\|x_i^+-z^*\|\right),\\
			G(Z^*)&\leq G(x^+)-\hat\Pi \left(\sum^N_{i=1}\|x_i^+-z^*\|\right),
		\end{split}\right.
	\end{equation}
	 where $Z^*=[(z^*)^\top,\cdots,(z^*)^\top]^\top\in \mathbb R^{nN}$, $ \tilde{\Pi}_i (\cdot)$ and $ \hat{\Pi} (\cdot)$ are class $\mathcal K$ functions.
	By summing up the first equation of \eqref{eq: aux_ineq}, the following equation is obtained,
	\begin{equation}\label{eq: auxiliary 1}\small
		\begin{split}
			&\sum_{i=1}^{N}\left(f_i(x_i^+)-f_i(z^*)\right)\\
			\leq&\sum_{i=1}^{N} ( \lambda_i+B_i(x_i^+-z))^\top(z^*-x_i^+)-\sum_{i=1}^{N}\tilde{\Pi}_i\left(\|x_i^+-z^*\|\right).
		\end{split}
	\end{equation}
	Similarly, from the second equation of \eqref{eq: aux_ineq},
	we have
	\begin{equation}\label{eq: auxiliary 2}
		\begin{split} &\sum_{i=1}^{N}\left(f_i(z^*)-f_i(x_i^+)\right)\\
			\leq&\sum_{i=1}^{N }(x_i^+-z^*)^\top \lambda_i
			-\hat\Pi \left(\sum^N_{i=1}\|x_i^+-z^*\|\right).
		\end{split}
	\end{equation}
	Combine Equations \eqref{eq: auxiliary 1} and \eqref{eq: auxiliary 2}, the following inequality can be obtained,
	\begin{equation}\label{eq: combine}
		\begin{split}
			&\sum_{i=1}^{N}( \lambda_i-\lambda_i^*+B_i( x_i^+-z))^\top(x_i^+-z^*)\\
			\leq& -\Pi\left(\sum^N_{i=1}\|x_i^+-z^*\|\right),
		\end{split}
	\end{equation}
	where 
	\begin{equation*}\small
		\begin{split}
			&\Pi\left(\sum^N_{i=1}\|x_i^+-z^*\|\right)\\
			=& \hat\Pi \left(\sum^N_{i=1}\|x_i^+-z^*\|\right)+ \sum_{i=1}^{N}\tilde{\Pi}_i\left(\|x_i^+-z^*\|\right)\in 
			\mathcal K.
		\end{split}
	\end{equation*}
	In Algorithm \ref{alg:BFGS ALADIN2}, since $\lambda_{i}^+=B_i(x_i^+-z^+)-g_i$ (Equation \eqref{eq: lam update}) and $g_i=B_i(z-x_i^+)-\lambda_i$ (Equation \eqref{eq: subgradient}), it is easy to show 
	\begin{equation}\label{eq: x update}
		x_i^+=\frac{B_i^{-1}}{2}(\lambda_i^+-\lambda_i)+\frac{1}{2}(z+z^+).
	\end{equation}
	On the other hand, from the KKT system of Problem \eqref{eq: consensus QP}, 
	\begin{equation}\label{eq: dual sum}\small
		\sum_{I=1}^{N}\lambda_i=0
	\end{equation}
	\eqref{eq: dual sum} is always guaranteed.
	
	By plugging \eqref{eq: x update} and \eqref{eq: dual sum} into the left hand side of \eqref{eq: combine},
	it can be shown that
	\begin{equation}
		\begin{split}\label{eq: proof1 end}\small
			\eqref{eq: combine}
			=&\frac{1}{4} \sum_{i=1}^{N} \left( -\|\lambda_i-\lambda_i^* \|_{B_i^{-1}}^2 + \|\lambda_i^+-\lambda_i^* \|_{B_i^{-1}}^2 \right)\\
			&+	\frac{1}{4}\left( \|z^+-z^*\|_{B_i}^2   - \|z-z^*\|_{B_i}^2\right)\\
			=& \frac{1}{4}\mathscr{L}(z^+,\lambda^+)-\frac{1}{4}\mathscr{L}(z,\lambda)\\
            \leq& -\Pi\left(\sum_{i=1}^N\|x_i^+-z^*\|\right).
		\end{split}
	\end{equation}
 
	This completes the proof.
	\hfill$\blacksquare$
	
	Note that, for smooth and strongly convex $f_i$s, we can establish the global linear convergence theory of Consensus ALADIN.
	
	\begin{theorem}\label{theorem 3 }
		Let the local objective  $f_i$s of Problem \eqref{eq: DC} be closed, proper, $m_f$ strongly convex
		and twice continuously differentiable such that the  gradients $g_i$s exist. Let the strong duality of Problem~\eqref{eq: DC} hold. Let matrices $B_i\succ 0$s be constant.
		Let $(z^*,\lambda^*)$ be the unique optimal primal and dual solution of Problem \eqref{eq: DC}. There exists a sufficiently small $\delta>0$ such that
		\begin{equation}\label{eq: dela}
			\delta \mathscr{L}(z^+,\lambda^+)\leq 4 m_f \sum_{i=1}^{N}\left\| x_i^+ -z^*\right\|^2,
		\end{equation}
		then the iteration of Algorithm \ref{alg:BFGS ALADIN2}  is linearly converging to $(z^*,\lambda^*)$.
	\end{theorem} 
	\textbf{Proof:}		Let $ f_i$s be twice continuously differentiable and $m_f$ strongly convex, then the following inequality satisfies 
	\begin{equation}\label{eq: linear rate}
		\begin{split}
			m_f \sum_{i=1}^{N}\left\| x_i^+ -z^*\right\|^2 
			\leq &\sum_{i=1}^{N} \left(x_i^+-z^* \right)^\top \left( g_i -g_i^* \right)\\
			\overset{\eqref{eq: combine}-\eqref{eq: proof1 end}}{=}& 	\frac{1}{4}\mathscr{L}(z,\lambda)-\frac{1}{4}\mathscr{L}(z^+,\lambda^+).
		\end{split}
	\end{equation}
	Combine the result of Equation \eqref{eq: linear rate} 
	and \eqref{eq: dela},
	we have	\[\mathscr{L}(z^+,\lambda^+)\leq \frac{1}{1+\delta}\mathscr{L}(z,\lambda).\]
	Later, a serial recurrence formula can be established:
	\begin{equation}\label{eq: global linear}
		\begin{split}
			\sum_{i=1}^{N}  \|z^k-z^*\|_{B_i}^2 &\leq \sum_{i=1}^{N}\left(\left\|\lambda_i^k-\lambda_i^*\right \|_{B_i^{-1}}^2 +   \left\|z^k-z^*\right\|_{B_i}^2\right)	\\
			&= \mathscr{L}(z^k,\lambda^k)\leq \left( \frac{1}{1+\delta} \right)^k 	\mathscr{L}(z^0,\lambda^0).
		\end{split}
	\end{equation} 
	Here $z^0,\lambda^0$ denote the initial primal and dual variables of the algorithm, respectively. Here $k$ represents the iteration index. 
	Equation \eqref{eq: global linear} shows the global linear convergence rate of Consensus ALADIN.  Theorem \ref{theorem 3 } is proved.
	\hfill$\blacksquare$
	
	Note that, by setting $B_i=\rho I$, the global convergence theory of Reduced Consensus ALADIN can also be established. 
	To the best of our knowledge, Theorem \ref{theorem 3 } provides the first global linear convergence theory for ALADIN-type algorithms in convex problems.

	\subsection{Local Convergence Analysis}\label{sec: local}
	The local convergence analysis of Consensus ALADIN for nonconvex problems is similar to \cite[Section 7]{houska2016augmented}.

	\begin{theorem}\label{them: local convergence}
		For Problem \eqref{eq: DC}, let $f_i$s be twice continuously differentiable. Let $x_i$s be updated as \eqref{eq: NLP}. Let the second order sufficient optimality condition (SOSC) be satisfied at a local minimizer $z^*$ of \eqref{eq: DC}.
		By initializing 
		$x_i$s and $z$ in a neighborhood of $z^*$,
		the iteration of Algorithm \ref{alg:BFGS ALADIN2} locally converges  with linear rate.
	\end{theorem}
	\textbf{Proof:} 
Let the minimizers of the decoupled problems \eqref{eq: NLP} be regular KKT points and lie within a neighborhood of the optimal solution $z^*$. Under these conditions, 
\eqref{eq: con_ALADIN_step1} is satisfied for
all the local problems \eqref{eq: NLP},
	\begin{equation}\label{eq: con_ALADIN_step1}
		\begin{split}
			\frac{\rho}{\sigma}\left\|z-z^{*}\right\|+\frac{1}{\sigma}\left\|\lambda_i-\lambda_i^*\right\|&\geq  \left\|x_i^{+}-z^*\right\|.
		\end{split}
	\end{equation}
	Here $\left\|\nabla^2f_i(x_i^+)+\rho I \right\|\geq \sigma>0$.
	
	For Consensus ALADIN, 
	let $\gamma$ denote the upper bound of distance between the Hessian approximation $B_i$ \eqref{eq: BFGS} and the real Hessian, i.e. $\left\|B_i-\nabla^2f_i(x_i^+)\right\|\leq \gamma$. From the standard SQP theory \cite[Chapter 18]{Nocedal2006}, 
	we have
	\begin{equation}\label{eq: step2-linear}
		\left\{
		\begin{split}
			N\|z^{+}-z^*\|&\leq \gamma\sum_{i=1}^{N}\|x_i^{+}-z^*\|,\\ \sum_{i=1}^{N}\|\lambda_i^{+}-\lambda_i^*\|&\leq \gamma\sum_{i=1}^{N}\|x_i^{+}-z^*\|.
		\end{split}\right.
	\end{equation} 
	Combine \eqref{eq: step2-linear} with \eqref{eq: con_ALADIN_step1},  the following inequality can be obtained.
	\begin{equation}\label{eq: linear convergence}
			\small
		\begin{split}
			&\left( 	\frac{\rho N}{\sigma}\|z^+-z^*\|+\frac{1}{\sigma}\sum_{i=1}^{N}\|\lambda_i^{+}-\lambda_i^*\|\right) \\
			\leq&\frac{\left( \rho +1\right)\gamma}{\sigma}\left( 	\frac{\rho N}{\sigma}\|z-z^{*}\|+\frac{1}{\sigma}\sum_{i=1}^{N}\|\lambda_i-\lambda_i^*\|\right).  
		\end{split}
	\end{equation}
 As long as the initial point of $x_i$s are sufficiently close to $z^*$, $\frac{\left( \rho +1\right)\gamma}{\sigma}<1$ is satisfied.
	This shows a local linear convergence of Consensus ALADIN. 
	
	This completes the proof.
	\hfill$\blacksquare$
	
	Importantly, \cite{Du2023B} proposed a globalization strategy for enforcing Consensus ALADIN globally converge to a local minimizer of non-convex DC.
	\section{Numerical Example}\label{sec: numerical}
	In this section, due to space limitations, we illustrate the numerical performance of the proposed algorithms only for non-convex DC. All algorithms are implemented using \texttt{Casadi-v3.5.5} with \texttt{IPOPT} \cite{Andersson2019}. 
	
	{\color{black}{By modifying a non-convex \emph{sensor allocation} problem from \cite{houska2016augmented}, we implement the folloing non-convex DC problem:}}
	\begin{equation}\label{eq: nonconvex problem}
		\small
		\begin{split}
			\min_{x_i,z,\forall i \in \mathcal I}&\quad \sum_{i=1}^{N}\left(\frac{1}{2}\left(\|x_i^\alpha- \zeta_i^\alpha\|^2+\|x_i^\beta- \zeta_i^\beta\|^2\right)+\right.\\
			&\quad\quad\;\;\left.\frac{1}{2}\sum_{j=1}^{10}\left(\left(x_{i}^\alpha[j]-x_{i}^\beta[j]\right)^2-\zeta_{i}^\sigma[j]\right)^2\right) \\ \quad\mathrm{s.t.}\quad&\quad x_i = z\  |\lambda_i, \;
		\end{split}
	\end{equation}
	where $x_i=[(x_i^\alpha)^\top,(x_i^\beta)^\top]^\top$, with $x_i^\alpha,x_i^\beta\in \mathbb R^{5}$ and $N=20$. 
	Here, $(\cdot)[j]$ denotes the $j$'s component of the given vector $(\cdot)$. All components of the measured data $(\zeta_i^\alpha,\zeta_i^\beta,\zeta_i^\sigma)$s are drawn from a Gaussian distribution $\mathcal N(0, 25) $ with proper dimension. In this setting, Problem \eqref{eq: nonconvex problem} has $210$ primal variables and  $200$ dual variables. In our implementation, the update of the local primal variables $x_i$s relies on \texttt{CasADi}. 
	Additionally, the hyper-parameter $\rho$ is set as $10^2$ for all algorithms. Note that all the initial values of primal and dual variables are set as zeros vectors in our implementation.
	
	\begin{figure}[H]
		\centering
		\includegraphics[width=0.38\textwidth,height=0.2\textheight]{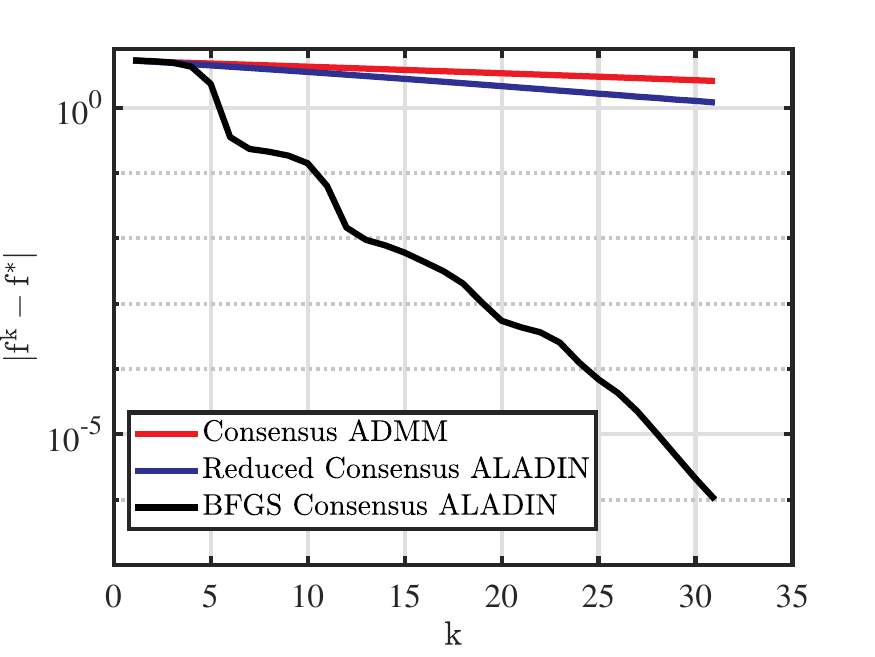}
		\caption{Numerical comparison on a non-convex case \eqref{eq: nonconvex problem}.}
		\label{fig:nonconvex}
	\end{figure}
	
	Figure  \ref{fig:nonconvex} shows the numerical convergence comparison among Consensus ADMM, Reduced Consensus ALADIN and BFGS Consensus ALADIN  for Problem  \eqref{eq: nonconvex problem} as a function of  iteration $k$. It can be observed that, even without second-order information, Reduced Consensus ALADIN still converges faster than Consensus ADMM. Additionally, BFGS Consensus ALADIN performs significantly superior performance compared to Reduced Consensus ALADIN for the given non-convex DC case. 
			
	If Problem \eqref{eq: nonconvex problem} is reformulated in the manner of \eqref{eq: DOPT_G2} and solved by Algorithm \ref{alg:ALADIN},
			the convergence behavior  is similar to that of BFGS Consensus ALADIN.  However, as explained  in Section \ref{sec: Preliminaries}, Algorithm \ref{alg:ALADIN} is not efficient  for solving \eqref{eq: nonconvex problem} due to high communication and computational overhead.
			Therefore,  the numerical performance of Algorithm \ref{alg:ALADIN} for \eqref{eq: nonconvex problem} is then not illustrated here. Interested readers are referred to \cite[Section 12]{ALADIN2023learning} for the implementation of Algorithm \ref{alg:ALADIN} on Problem \eqref{eq: DOPT_G2}.

	\section{Conclusion}\label{sec: summary}
	This paper introduces a novel family of algorithms, named Consensus ALADIN, designed to efficiently solve distributed consensus optimization problems of the form \eqref{eq: DC}. Building on the structure of Consensus ALADIN, we propose BFGS consensus ALADIN,  a communication-and-computation-efficient variant. 
 Furthermore, we introduce a more computationally-efficient algorithm, called Reduced  Consensus ALADIN, based on BFGS Consensus ALADIN.  We establish convergence theory for  Consensus ALADIN and conduct numerical experiments to demonstrate its practical effectiveness.
	
	
	\section*{Acknowledgement}
The authors wish thank Xiaohua Zhou and Shijie Zhu from ShanghaiTech University for helpful discussions.

	\appendix
	
	\section{Comparison Between Reduced Consensus ALADIN and Consensus ADMM}\label{sec: cadmm}
	In the standard Consensus ADMM framework, there are two different cases: a) first update the dual then aggregate (update the global variable) \cite{zhou2023federated}, b) first aggregate  then update the dual \cite[Chapter 7]{boyd2011distributed}. 
	In order to distinguish the Reduced Consensus ALADIN \eqref{eq: RCA} from the two variant of  Consensus ADMM, we use superscripts on key variables, such as $\lambda_i$s and $z$,  to show the difference.

	\subsection{First Update the Dual then Aggregate} 
	By using  the dual variable $\lambda_i$,
	the corresponding Lagrangian function of \eqref{eq: DC} can be expressed as
	\begin{equation}\label{eq: ADMM Lagrangian}
		\begin{split}
			\mathcal{L}(x, z, \lambda)=&\sum_{i=1}^{N}\left(f_i(x_i)
			+\lambda_{i}^\top (x_i-z) + \frac{\rho}{2}\|x_i-z\|^2\right),
		\end{split}
	\end{equation}
	where $\rho$ is a given positive penalty parameter.
	{\color{black}{From \eqref{eq: ADMM Lagrangian}, the main steps of  updating the local primal and dual variables with Consensus ADMM \cite{zhou2023federated} can be summarized as the following equation, 
	\begin{equation}\label{eq: ADMM1}
		\small
		\left\{
		\begin{split}
			&{x_i}^+=\mathop{\arg\min}_{x_i}f_i(x_i)+\left(\lambda_i^{\text{ADMM1}}\right)^\top  x_i+\frac{\rho}{2}\left\|x_i-z^\text{ADMM1}\right\|^2,\\
			&\left(\lambda_i^{\text{ADMM1}}\right)^+=\left(\lambda_i^{\text{ADMM1}}\right)+\rho\left(x_i^+-z^\text{ADMM1}\right),\\[0.8em]
			&\left(z^\text{ADMM1}\right)^{+}=	\frac{1}{N} \sum_{i=1}^{N}\left(x_i^++ \frac{\left(\lambda_i^{\text{ADMM1}}\right)^+}{\rho}\right).\\
		\end{split}\right.
	\end{equation}
	From the expression of the gradients \eqref{eq: subgradient} in Consensus ALADIN \eqref{eq: RCA},  one may find 
	\begin{equation}\label{eq: lam-g}
		\left(\lambda_i^{\text{ADMM1}}\right)^+ = - g_i. 
	\end{equation}

	It can be noticed that the framework of Reduced Consensus ALADIN \eqref{eq: RCA} is very similar to this order of \eqref{eq: ADMM1}. 
	By setting $B_i=\rho I$, \eqref{eq: consensus QP} boils down to 
	\begin{equation}\label{eq: rhoALADIN}
		\begin{split}
			&\left(z^\text{ALADIN}\right)^{+}\\
			=&\left\{
			\begin{split}
				\mathop{\arg\min}_{ \Delta x_i, z,\forall i \in \mathcal I}& \quad \mathop{\sum}_{i=1}^{N} \left(\frac{\rho}{2}\Delta x_i^\top \Delta x_i+g_i^\top \Delta x_i\right) \\
				\mathrm{s.t.} \quad&\qquad\Delta x_i+x_i^+=z\; |\lambda_{i}^{\text{ALADIN}}
			\end{split}\right\}.
		\end{split}
	\end{equation}
	It is equivalent to the same operation as that of Consensus ADMM framework \eqref{eq: rhoADMM} (if we ignore the auxiliary variables $\Delta x_i$s)
	\begin{equation}\label{eq: rhoADMM}
		\begin{split}
			&\left(z^\text{ADMM1}\right)^{+}\\
			=&	\mathop{\arg\min}_{z} \; \mathop{\sum}_{i=1}^{N} \left(\frac{\rho}{2}\| x_i^+-z\|^2-\left(\lambda_i^{\text{ADMM1}}\right)^\top z\right).
		\end{split}
	\end{equation}
	In both ways of updating the global variable $z$, Equation \eqref{eq: rhoALADIN} and \eqref{eq: rhoADMM}, have the same result as Equation \eqref{eq: reduced QP2}.  
	However, the dual update in Reduced Consensus ALADIN is  different from \eqref{eq: lam-g} which is shown as the following equation,
	\begin{equation}\label{eq: ALADIN dual}
		\lambda_{i}^{\text{ALADIN}}=\rho\left(x_i^+-\left(z^\text{ALADIN}\right)^{+}\right)-g_i .
	\end{equation}
	\vspace{0.11cm}
	Importantly,  in the Consensus ADMM iteration \eqref{eq: ADMM1}, $\sum_{i=1}^N \left(\lambda_i^\text{ADMM1}\right)^* =0$
	is guaranteed only at the optimal point.
	On the opposite, with  Reduced Consensus ALADIN \eqref{eq: RCA}, $\sum_{i=1}^N \lambda_i^{\text{ALADIN}} =0$
	is guaranteed in each iteration.
	Since the first version of Consensus ADMM can not bring the latest dual update back to each agent, 
	the Reduced Consensus ALADIN framework can be interpreted as a more efficient way for using the consensus dual information.

	\subsection{First Aggregate then Update the Dual}\label{sec: admm2}
	
	As introduced in \cite[Chapter 7]{boyd2011distributed}, Consensus ADMM can be interpreted as the following equation,
	\begin{equation}\label{eq: ADMM22}
		\small
		\left\{
		\begin{split}
			&{x_i}^+=\mathop{\arg\min}_{x_i}f_i(x_i)+\left(\lambda_i^{\text{ADMM2}}\right)^\top  x_i+\frac{\rho}{2}\left\|x_i-z^{\text{ADMM2}}\right\|^2,\\
			&\left(z^\text{ADMM2}\right)^{+}=	\frac{1}{N} \sum_{i=1}^{N}\left(x_i^++ \frac{\lambda_i^{\text{ADMM2}}}{\rho}\right),\\
			&\left(\lambda_i^{\text{ADMM2}}\right)^+=\left(\lambda_i^{\text{ADMM2}}\right)+\rho\left(x_i^+-\left(z^\text{ADMM2}\right)^{+}\right).
		\end{split}\right.
	\end{equation}
	With \eqref{eq: ADMM22}, Consensus ADMM  can also converge for convex problems with guarantees.
	In this form, the update of $\lambda_i^{\text{ADMM2}}$s has the same property as Reduced Consensus ALADIN that  guarantees 
	\begin{equation}\label{eq: ADMM2}
		\sum_{i=1}^N \lambda_i^{\text{ADMM2}} =0
	\end{equation}
	in each iteration. In this way, the dual variables can also carry sensitivity information of the  gap between the latest global variable $\left(z^\text{ADMM2}\right)^{+}$ and the local variables $x_i^+$. 
	However, the second version of Consensus ADMM can not upload the latest local dual \eqref{eq: lam-g} back to the master because of \eqref{eq: ADMM2}. This shows that Reduced Consensus ALADIN \eqref{eq: RCA} is more efficient for global variable aggregation than \eqref{eq: ADMM22}.  
	\bibliographystyle{ieeetr}
	\bibliography{paper}
	
	
	\balance

\end{document}